\documentclass[10pt,aps,prl,superscriptaddress,showpacs,preprint]{revtex4-1}
\usepackage{graphicx}

\begin{document}

\title{Trapping Surface Electrons on Graphene Layers and Islands}

\author{D. Niesner}
\author{Th. Fauster}

\affiliation{Lehrstuhl f\"ur Festk\"orperphysik, Universit\"at 
Erlangen-N\"urnberg, D-91058 Erlangen, Germany}

\author{J. I. Dadap}
\author{N. Zaki}
\author{K. R. Knox}
\author{P.-C. Yeh}
\author{R. Bhandari}
\author{R. M. Osgood}

\affiliation{Columbia University, New York, New York 10027, USA}

\author{M. Petrovi\'{c}}
\author{M. Kralj}

\affiliation{Institut za fiziku, Bijeni\v{c}ka 46, HR-10000 Zagreb, 
Croatia}

\begin{abstract}
We report the use of time- and angle-resolved two-photon photoemission
to map the bound, unoccupied electronic structure of the weakly coupled
graphene/Ir(111) system.  The energy, dispersion, and lifetime of the
lowest three image-potential states are measured.  In addition, the weak
interaction between Ir and graphene permits observation of resonant
transitions from an unquenched Shockley-type surface state of the Ir 
substrate to graphene/Ir image-potential states.  The
image-potential-state lifetimes are comparable to those of mid-gap clean
metal surfaces.  Evidence of localization of the excited electrons on
single-atom-layer graphene islands is provided by coverage-dependent
measurements.
\end{abstract}


\maketitle

Graphene on metal surfaces are a materials system of enormous
fundamental and applied interest.  The graphene/metal interface is
encountered in the rapidly expanding technological system of CVD
graphene on Cu foil, in the structurally precise monolayer epitaxial
systems of graphene on single-crystal Ru, Ir, or Ni, and finally in the
metal contacts of graphene field-effect transistors or other devices.
Questions then arise on the electronic structure of graphene on metal
surfaces and in fact several recent studies have addressed questions
such as the role of lattice mismatch on band structures.  Most studies
of the electronic structure of graphene have focused on the band
structure in the vicinity of its K point, near the Fermi edge.  Further
there has been a paucity of measurements about its unoccupied electronic
structure and the dynamics of strongly excited electrons.
Image-potential states offer one important approach to probe the excited
state manifold and are known to vary with interfacial quality,
dielectric properties, and electronic structure.  In graphene the large
band gap at the $\Gamma$ point results in Bragg reflection from the
crystal within a certain range of energy and momentum.  In fact a recent
theoretical study has shown the existence of a dual Rydberg-like series
of even and odd symmetry image-potential states in a single
free-standing sheet of graphene \cite{SIL2010}.  Image-potential states
on graphene may experience different dynamic constraints.  For example,
the different phase space for decay in two dimensions compared to three
dimensions may affect the lifetimes for electrons trapped in
image-potential states on graphene.  In fact, more generally the
response of the image electron to the composite dielectric/metal systems
is itself of basic physics interest.

\begin{figure}
\includegraphics[width=.67\columnwidth]{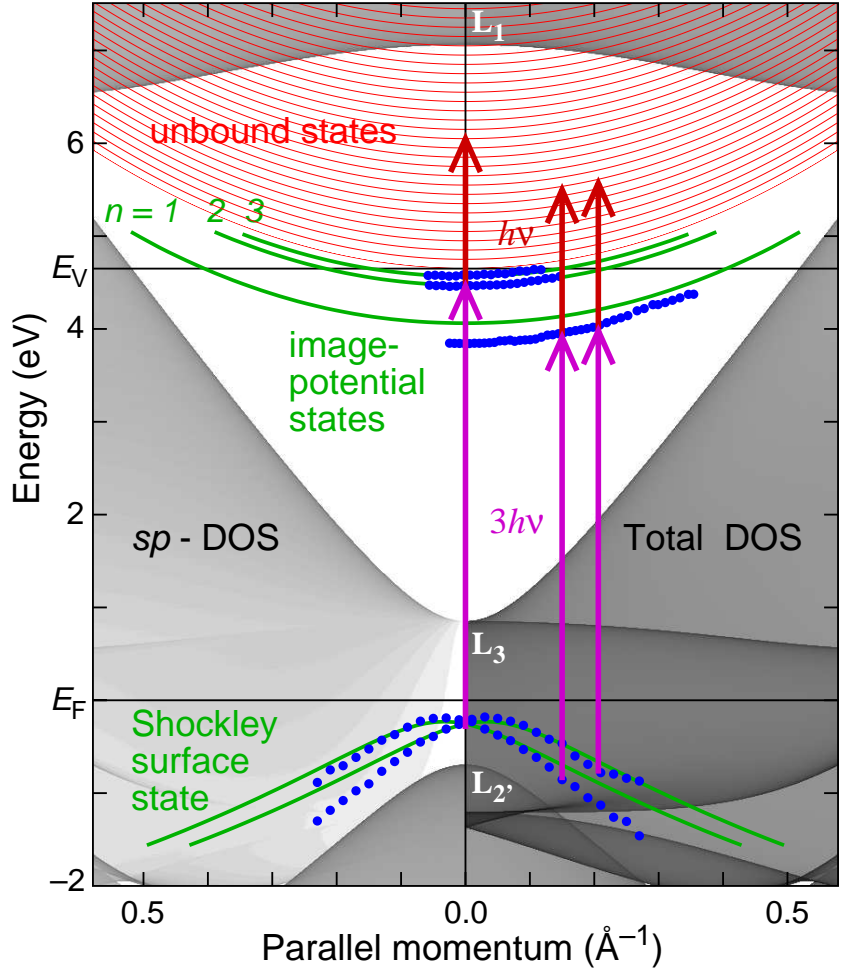}

\caption{(Color online).  
Arrows indicate 2PPE transitions between surface and image-potential
states.  The experimental results (dots) are compared to
calculations (lines).  The projected bulk-band structure of
Ir(111) along the $\Gamma$K direction is shaded according to the total
and $sp$-density of states (DOS) at the right and left, respectively.}
\label{bs} \end{figure}

In this Letter, we investigate the uncharted region of the bound,
unoccupied electronic structure of epitaxial graphene grown on Ir(111)
in the vicinity of the graphene $\Gamma$ point; our measurements are
made via the image-potential states using angle- and time-resolved 
two-photon photoemission (2PPE) as indicated by arrows 
in Fig.~\ref{bs}.  This system was chosen for several reasons:
First, because of the weak coupling in the graphene/Ir(111) system, the
electronic structure of the graphene overlayer is nearly intact, with
sharp Dirac dispersion characteristics \cite{PLE2009}.  In addition, the
moir\'e corrugation of the epitaxial graphene on Ir(111) has been found
to be only $0.35\pm0.10$~{\AA} based on atomic force microscopy
measurements \cite{SUN2011} indicating a smooth epitaxial graphene
surface.  Second, the molecular-based growth is well characterized and
saturates at precisely one monolayer (ML) of epitaxial graphene
\cite{GAS2009}.  Our results show that image-potential states may be
excited from the Ir/graphene interfacial region and have binding
energies and lifetimes comparable to those of mid-gap clean metal
surfaces.  In addition, spectral measurements of binding energy versus
coverage show clearly that at low graphene coverage, image-potential
electrons are trapped on graphene islands by surface work function
differences between the metal and graphene regions, an observation of
high importance for understanding of transport at graphene-metal
interfaces \cite{XIA2011}.

Our choice of two-photon photoemission is the result of its high
temporal and energy resolution.  Other experimental observations of
image-potential states have used scanning tunneling spectroscopy, i.~e.,
graphene on SiC \cite{BOS2010} and on Ru(0001) \cite{BOR2010}.  This
technique, however, measures the image-potential series in the presence
of strongly distorting electric field between tip and sample and without
time-resolved possibilities.

The experiments were conducted using monochromatic and bichromatic 2PPE,
and angle-resolved photoemission (ARPES).  Details of the monochromatic 
2PPE setup at Columbia which was used in the photon energy range of
$3.8<h\nu<4.9$~eV are given in Ref.~\cite{HAO2010}.  Bichromatic and
time-resolved 2PPE measurements were performed in Erlangen using
pump-probe methods with the third harmonic (UV) and the fundamental (IR,
$1.51<h\nu<1.62$~eV) as described in Ref.~\cite{Boger05njp}.
Additionally, occupied-state ARPES measurements were performed at APE
(ELETTRA) using a photon energy of 55~eV with an energy resolution of
20~meV\@.  The resolution of the 2PPE experiments was 40~meV\@.  The
base pressure in all three ultrahigh vacuum (UHV) systems was better
than $1\times10^{-8}$~Pa.  All measurements used $p$-polarized beams.

Graphene was prepared by cycles of temperature programmed growth, TPG
(room temperature ethene exposure $6\times10^{-6}$~Pa for 60~s and
flashed to $\approx1450$~K), followed by a chemical vapor deposition run
($6\times10^{-6}$~Pa of ethene for 300~s at 1150~K), to form exactly one
graphene monolayer \cite{GAS2009}.  Growth was monitored by low-energy
electron diffraction (LEED) after each cycle, which showed the
development of the characteristic moir\'e pattern of uniformly oriented
graphene \cite{HAT2011}, as graphene coverage varied from 0
to 1~ML; LEED patterns (not shown) revealed these patterns clearly.

\begin{figure}
\includegraphics[width=\columnwidth]{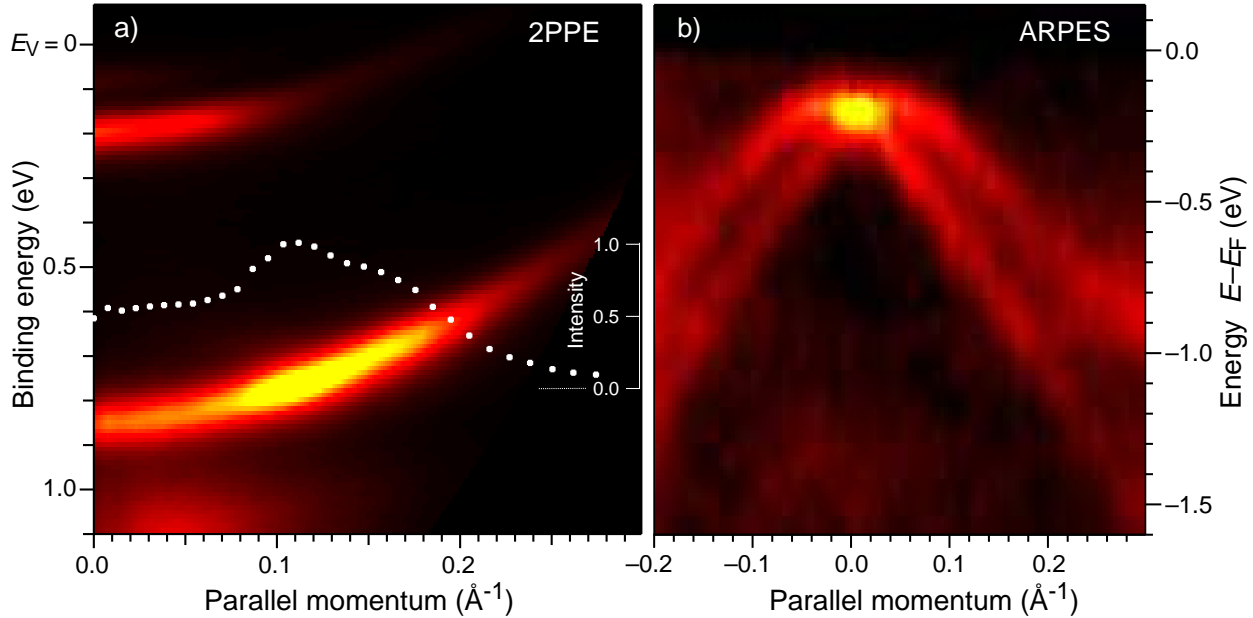}

\caption{(Color online).  a) Intensity map of the 2PPE signal recorded
with photon energy $h\nu=1.59$~eV for 1~ML graphene on Ir(111).  Points
represent the intensity of the lowest $n=1$ band.  b) ARPES map showing
initial states for $h\nu=55$~eV\@.}
\label{maps} 
\end{figure}

Figure~\ref{maps}a shows the measured 2PPE intensity obtained at
1~ML and for $h\nu=1.59$~eV along the $\Gamma$K direction.
Three unoccupied bands are observed.  The pumping process could be
deduced from its photon-energy dependence, thus in bichromatic case, all
peak positions shifted linearly with IR photon energy, indicating that
the process involves pumping by a UV photon and photoemission by an IR
photon \cite{HOF1997}.  All 2PPE features vanished when the IR beam was
switched to $s$-polarization, indicating the expected symmetry for
image-potential states.  The effective masses of all three states are
$0.9\pm0.1\,m_{e}$.  The binding energies of the three states with
respect to the vacuum level are given in Table~\ref{tab1}.  The measured
energies and effective masses are close to the free electron mass and
fit well to a Rydberg-like-series of image-potential states with a
nonvanishing quantum defect \cite{EP1978}.

Figure~\ref{maps}a shows that the $n=1$ band is most intense for
parallel momenta $k_\|$ between 0.08 and 0.17~\AA\textsuperscript{-1}
{[}cf.\ points in Fig.~\ref{maps}a{]}, with the intensity typically decreasing
monotonically with increasing $k_\|$ \cite{GUD2007}.  Direct transitions
from initial surface bands can lead to intensity resonances
\cite{HAO2010}.  In order to identify possible initial states for 2PPE,
we have performed ARPES measurements of graphene on Ir(111).  The ARPES
data in Fig.~\ref{maps}b show two parabolic-like dispersions with a
downward curvature.  The two branches are shifted from $k_\|=0$ by
$\pm0.033\pm0.001$~\AA\textsuperscript{-1} and have a maximum energy of
$-0.19\pm0.01$~eV\@.  Similar results were also obtained with the fourth
harmonic (6.2~eV) in the 2PPE setup.  Rashba-type splittings of similar
magnitude are found in other systems, e.g., a Bi/Ag(111) surface alloy
\cite{AST2007}.  These bands are also observed on clean Ir(111)
\cite{VEEN1980,PLE2010} indicating that this surface feature is inherent
to the clean metal surface.  The surface-state energy reported
for clean and graphene-covered surfaces differs by about 0.2 eV, an
effect, which is consistent with a charge transfer between substrate and
overlayer and which can shift graphene \cite{PLE2010} or iridium states
\cite{KRA2011,SUB2011}.

The initial band dispersion is plotted together with the measured
dispersion of the image-potential states in Fig.~\ref{bs} (blue dots).
The arrows connecting initial states to the $n=1$ image-potential band
are at slightly larger $k_\|$ values than the enhanced intensity in
Fig.~\ref{maps}a.  In the absence of resonances the 2PPE intensity along
image-potential bands decreases continuously with increasing parallel
momentum \cite{GUD2007}.  In the present case due to the finite energy
and angle resolution the intensity maxima are shifted to lower $k_\|$
values compared to the position found in the dispersion analysis.  The
additional resonance into the $n=2$ band (see Fig.~\ref{bs}) can be
inferred from the similar intensity as for the $n=1$ state at $k_\|=0$
in Fig.~\ref{maps}a and is confirmed by photon-energy-dependent data
presented in the supplemental material \cite{suppmat}.

\begin{table}
\caption{Experimental and calculated binding energies 
and lifetimes for image-potential states on graphene/Ir(111).}
\label{tab1} 
\begin{ruledtabular}
\begin{tabular}{cccc}
$n$ & $E^\mathrm{exp}_n$ (eV)& $E^\mathrm{calc}_n$ (eV) & $\tau$ (fs) \\ 
\hline 1 & $0.83\pm0.02$ & 0.59 & $35\pm3$ \\
2 & $0.19\pm0.02$ & 0.18 & $114\pm6$ \\
3 & $0.09\pm0.02$ & 0.08 & $270\pm12$ \\
\end{tabular}
\end{ruledtabular}
\end{table}

In order to understand the character of the initial state, we calculated
the projected bulk-band structure of Ir(111) using a non-relativistic
parameterized tight-binding scheme \cite{PAP1986}.  Figure~\ref{bs}
shows this projected structure along the $\Gamma$K direction, at the
right.  The shading represents the one-dimensional density of states
(1D-DOS).  The left part of Fig.~\ref{bs} shows the 1D-DOS of bands
according to their $sp$-character.  For $k_\|=0$, the lower edge of the
$sp$-band gap is at $-0.7$~eV, which corresponds to the $L_{2'}$ point.
The band edge of the total projected bulk-band structure disperses
upward from the $L_{3}$ point around $+0.8$~eV and picks up
$sp$-contributions.  On the other hand the lower $sp$-band edge shows a
downward dispersion.  The energy of the Shockley surface state was
calculated using the $sp$-band edges within a scattering model
\cite{FAU1994}.  The calculations used the experimental work function of
4.65~eV for graphene on Ir(111).  The calculated bands were shifted by
$\pm0.033$~\AA\textsuperscript{-1} to account for the experimentally
observed Rashba splitting and are drawn as green lines in Fig.~\ref{bs}
in the region below the Fermi energy.  The experimentally extracted
dispersion shown by dots agrees well with the calculation.  The
Shockley-type surface state is apparently not quenched by the graphene
layer at a distance of 3.4~\AA{} \cite{BUS2011}, because its probability
density is concentrated at the Ir(111) surface.

The scattering model was also used to calculate the energies of the
image-potential-band series \cite{FAU1994}.  The calculated binding
energies, given in Table~\ref{tab1}, are approximately those expected
for states located near midgap (see Fig.~\ref{bs}).  However, the
calculated $n=1$ binding energy is significantly smaller than the
experimental value.  This discrepancy is due to the fact that the
scattering model calculation neglects the round-trip phase shift
$2\phi_{gr}$ of the graphene layer.  Using the expressions for the phase
shift at the substrate and the image-potential barrier \cite{SMI1985} we
obtain $\phi_{C}=0.63\pi$ and $\phi_{B}=1.02\pi$, respectively.  The
total phase shift for the $n=1$ state is $2\pi$, from which we obtain
$\phi_{gr}=0.18\pi$.  Note that such a small phase shift leads to a
significant change in binding energy from 0.59~eV to 0.83~eV (see
Table~\ref{tab1}).

\begin{figure}
\includegraphics[width=\columnwidth]{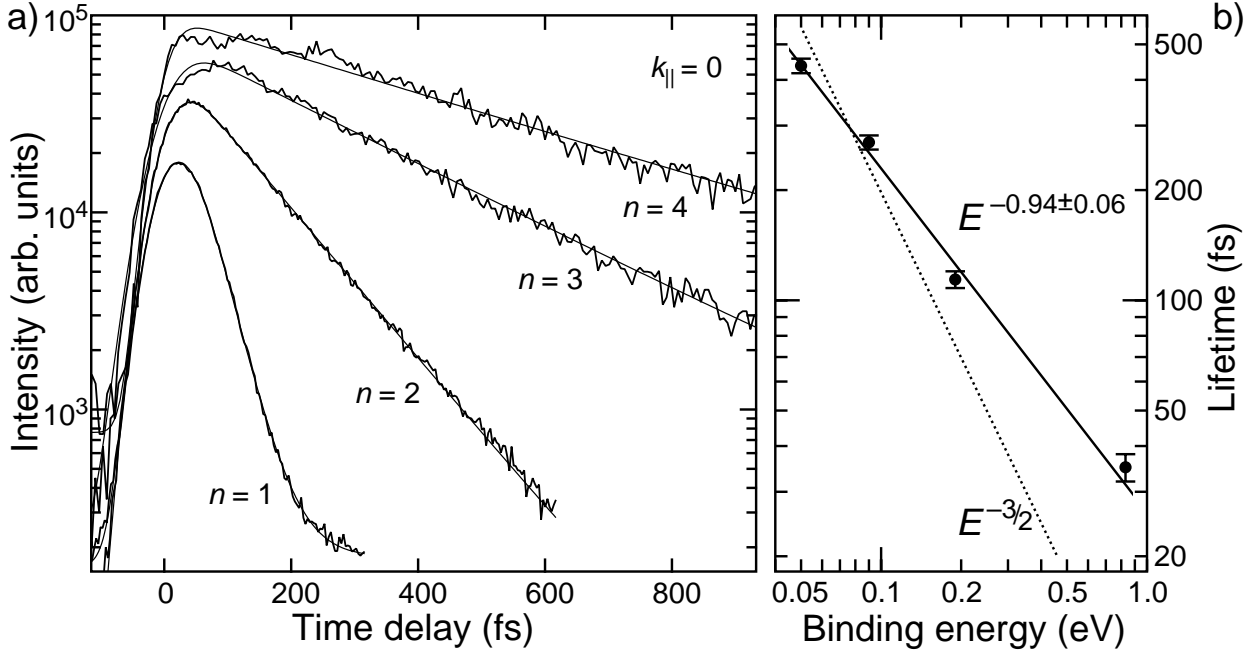}

\caption{a) Time-resolved measurements of the
image-potential-state series at $k_\|=0$. b) Lifetimes as function of 
binding energy compared to various power-law dependencies (see text).} 
\label{tr} \end{figure}

The time-resolved spectra of the image-potential states were also
measured and are shown in Fig.~\ref{tr}a.  As summarized in
Table~\ref{tab1}, lifetimes of tens to hundreds of femtoseconds are
obtained. These are comparable to values obtained for Cu(100) 
with a similar midgap image-potential state position and hence bulk
evanescent decay length in the metal crystal \cite{HOF1997}.  Note, as
an aside, that the curve measured at the energy of the $n=4$
image-potential state in Fig.~\ref{tr}a shows weak quantum beats
\cite{HOF1997} for delay times $<300$~fs.  The data in Table~\ref{tab1}
show that lifetimes vary with binding energy approximately $\propto
E^{-1}$ (solid line in Fig.~\ref{tr}b).  The asymptotic, classical
$\tau\propto E^{-3/2}$ behavior \cite{FAU2000} (dashed line in
Fig.~\ref{tr}b) is not reached for $n<4$.  Similar behavior has been
found on copper surfaces \cite{ssr}.

An important issue for carrier movement at graphene/metal interfaces is
the degree of lateral confinement.  This confinement can be examined at
low graphene coverage, obtained via a small number of sequential TPG
cycles.  From previous studies, it is known that one TPG cycle covers a
fraction of about 20\% of uncovered Ir surface \cite{KRA2011}.  After
one TPG cycle the typical island size is $(35~\mathrm{nm})^{2}$ and
after the second cycle of the order of $(100~\mathrm{nm})^{2}$
\cite{KRA2011}.

\begin{figure}
\includegraphics[width=.7\columnwidth]{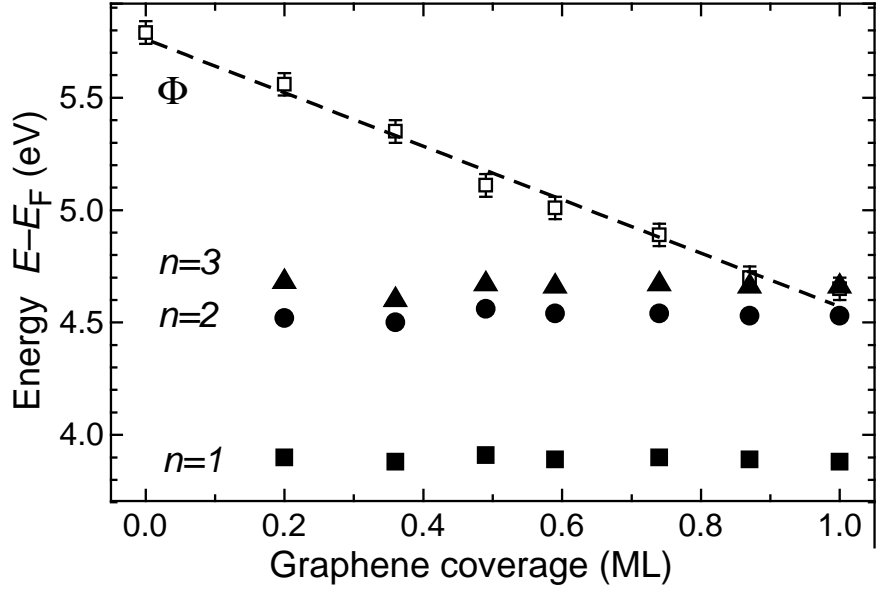}

\caption{Sample work function (open symbols) and
image-potential states $n=1$, 2, and 3 binding energies (filled symbols)
as a function of graphene coverage.  Dashed line represents a linear fit
for the work-function change.}
\label{cov} 
\end{figure}

As has been shown in earlier work \cite{FAU1993a}, the average and local
work function play an important role in interfacial electron
localization.  Thus the average work function $\Phi$ was measured via
monochromatic 2PPE and the expression $\Phi=2h\nu-\Delta E$, i.e.,
where $h\nu$ is the photon energy and $\Delta E$ is the difference
between the Fermi level cutoff and the low-energy cutoff. 
Figure~\ref{cov} displays the work function (open symbols)
as a function of graphene coverage.  The work function decreases
approximately linearly from a value $5.79\pm0.10$~eV to $4.65\pm0.10$~eV
from Ir(111) to 1 ML graphene.  Reported values of the work function for
Ir(111) are 5.76 and 5.79~eV \cite{Ir111WF}.  The work function of the
graphene-covered surface on Ir(111) is between the values for Pt(111) of
4.87~eV and free-standing graphene of 4.48~eV \cite{GIO2008}, which is
consistent with the weak bonding between the Ir(111) and the graphene
overlayer and a $p$-doping of the graphene \cite{PLE2009}.  The linear
decrease of the work function is known for other systems and is due to
the averaging over substrate and overlayer islands 
\cite{FAU1993a}.

Image-potential states were observed at all coverages reported here
using 2PPE\@.  However, for the clean surface or uncovered substrate 
areas the available photon energies were not sufficient to populate
image-potential states due to the large work function of Ir(111).  The
image-potential-state energies, measured relative to the Fermi level,
are shown in Fig.~\ref{cov}.  The energies are generally constant over
the coverage range from 0.2 to 1~ML with the intensity increasing
monotonically with coverage.  Note that the graphene Dirac cone at the K
point has been clearly observed for more than three TPG cycles or
0.5~ML graphene coverage \cite{KRA2011}.  The constant energy of the
image-potential series as a function of coverage in Fig.~\ref{cov} is a
direct result of the localization of the electrons on the graphene
islands \cite{FAU1993a}.  The electrons respond to the local work
function if the average island dimensions are larger than the typical
distance of the probability density maximum which is of the order of
nanometers for the lowest $n$ image-potential states.  Note that the
localization on the graphene islands is facilitated by the large work
function difference between the graphene layers and the Ir(111)
substrate.  For small graphene islands, an energy shift proportional to
$d^{-2}$, where $d$ denotes the characteristic island size, is expected
due to the lateral localization of the electron in a two-dimensional
quantum well \cite{FAU1993b}.  However, these shifts would be $<1$~meV
for the island sizes expected for the current preparation conditions
\cite{KRA2011}.

In summary, we have observed and measured the properties of
image-potential states on a graphene monolayer on Ir(111).  The binding
energy of the $n=1$ image-potential state is 40\% larger than expected
from the position of the graphene vacuum level relative to the Ir(111)
band gap.  There is no prominent indication of a second main series of
image-potential states as predicted for free-standing graphene
\cite{SIL2010}.  Apparently, the underlying metal substrate breaks the
mirror symmetry of the graphene layer and the state of odd symmetry
shifts up in energy as has been calculated for graphene on Ru(0001)
\cite{BOR2010}.  In addition, the image-potential states can be excited
efficiently from a downward dispersing Shockley surface state in the
$sp$-band gap of the Ir(111) band structure indicating a sizable overlap
of the wave functions of these states located at the substrate interface 
and graphene surface, respectively. The measured lifetimes of the
image-potential states are comparable to similar clean metal surfaces.
Recently, similar results have also been obtained for graphene on
Pt(111) \cite{Pt111}.  Apparently, the evanescent coupling of the
image-potential-state wavefunctions to the underlying electronic states
of the Ir(111) bulk and surface states is not altered by the graphene
layer.  Three-dimensional localization of electrons on graphene islands
has been observed for submonolayer coverages obtained by individual TPG
cycles.  However, even for the smallest island size, no energy shift due
to localization was observed within the experimental uncertainty.
Further development is needed to prepare well-ordered graphene islands
with controlled lateral extension.  A different approach would be to
exploit the moir\'e pattern on more corrugated graphene layers
\cite{BOR2010}.

\begin{acknowledgments}
The work at Columbia University was supported under the US Department of
Energy Contract No.\ DE-FG 02-04-ER-46157.  The Zagreb group
acknowledges support by the MZOS through project No.\ 035-0352828-2840
and MZOS-NSF support through contract No.\ 1/2009.  M.~K.\ thanks the
Alexander von Humboldt Foundation for a research fellowship.  We
acknowledge helpful discussions with Branko Gumhalter.
\end{acknowledgments}

\end{document}